\documentclass[%
 reprint,
 amsmath,amssymb,
 aps,
longbibliography]{revtex4-1}

\usepackage{graphicx}
\usepackage{dcolumn}
\usepackage{bm}
\usepackage{lineno}
\usepackage{subfig}

\begin{document}


\title{Tile Calorimeter Upgrade Program for the Luminosity Increasing at the LHC}%

\author{A. S. Cerqueira on Behalf of the ATLAS Tile Calorimeter System}
\affiliation{Electrical Engineering Department, Federal University of Juiz de Fora}
\collaboration{ATLAS Collaboration}


\date{\today}


\begin{abstract}
The Tile Calorimeter (TileCal) is the central hadronic calorimeter of the ATLAS experiment at the Large Hadron Collider
(LHC). 
The LHC is scheduled to undergo a major upgrade, in 2022, for the High Luminosity LHC (HL-LHC).
The ATLAS upgrade program for high luminosity is split into three phases: Phase-0 occurred during $2013-2014$, and prepared the LHC for Run~2; Phase-I, foreseen for 2019, will prepare the LHC for Run~3, whereafter the peak luminosity reaches $2-3 \times 10^{34}$~cm$^{2}s^{-1}$; finally, Phase-II, which is foreseen for 2024, will prepare the collider for the HL-LHC operation ($5-7 \times 10^{34}$~cm$^{2}s^{-1}$).
The TileCal main activities for Phase-0 were the installation of the new low voltage power supplies and the activation of the TileCal third layer signal for assisting the muon trigger at $1.0<|\eta|<1.3$ (TileMuon Project). In Phase-II, a major upgrade in the TileCal readout electronics is planned. Except for the photomultipliers tubes (PMTs), most of the on- and off-detector electronics will be replaced, with the aim of digitizing all PMT pulses at the front-end level. This work describes the TileCal upgrade activities, focusing on the TileMuon Project and the new on-detector electronics.
\end{abstract}

\maketitle


\section{Introduction}
The ATLAS experiment \cite{ATLAS} is one of the two multi-purpose experiments at the Large Hadron Collider (LHC), which collected approximately 26~fb$^{-1}$
of pp collisions during the LHC run~1 (2009-2012) with a center of mass energy of 7 and 8~TeV. The ATLAS calorimeters play an important role in the experiment, absorbing and sampling the energy of incoming electromagnetic and hadronic particles. The electromagnetic lead/liquid argon (LAr) calorimeter followed by the hadronic Tile calorimeter (TileCal) cover the central region of the ATLAS experiment up to a pseudorapidity of $|\eta|<1.7$, other LAr based calorimeters span across the forward regions, up to $|\eta|<4.9$. Together with the electromagnetic calorimeter, TileCal provides precise measurements of hadrons, jet, taus and missing transverse momentum.

\begin{figure}
\begin{center}
\includegraphics[width=6cm]{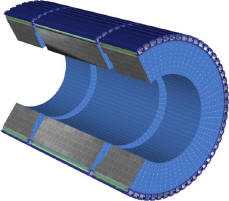}
\end{center}
\caption{\label{tilecal} Illustration of the Tile Calorimeter.}
\end{figure}

TileCal \cite{tile} is a sampling calorimeter composed of steel plates (tile shape) as absorber material interleaved with plastic scintillating plates as active material. It is divided in a central barrel (covering $|\eta|<1.0$) and two extended barrels (covering $0.8<|\eta|<1.7$), where each part is formed by 64 modules in order to complete the entire cylinder (see FIG.~\ref{tilecal}).

\begin{figure}
\begin{center}
\includegraphics[width=8cm]{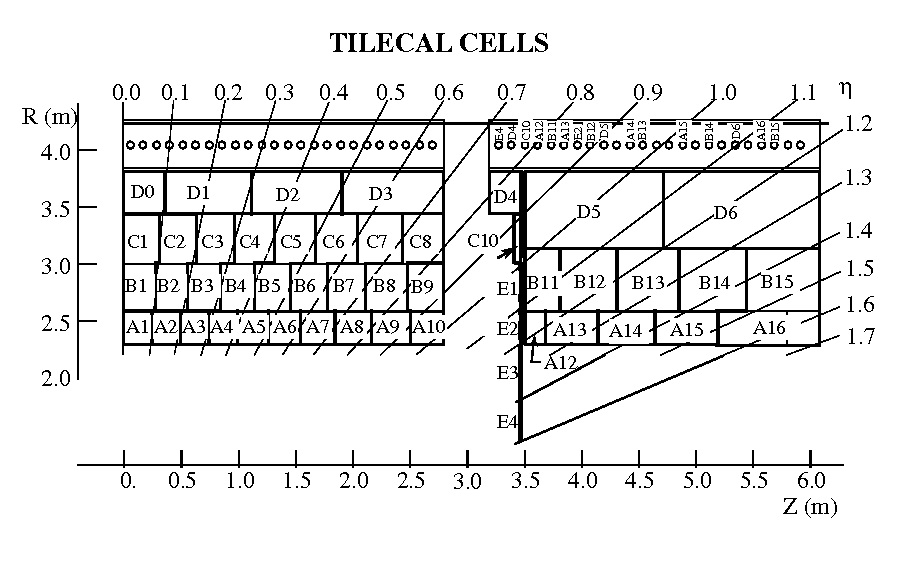}
\end{center}
\caption{\label{tilecalcells} TileCal cells definition.}
\end{figure}

When high energy particles interact with the steel, showers of lower energy particles are created, which in turn produce
light when passing through the scintillating tiles. The light is transmitted through wavelength shifting fibers to
photomultipliers (PMTs), which convert the light into electrical signals. Adjacent tiles and their WLS fibers are
grouped together to form TileCal cells. For each cell the fibers are read-out from two sides by two PMTs. Thus, each cell is read out via two different electrical signal paths. The central barrel modules are divided in up to 45
cells each, while the extended barrels modules are divided in 14 cells. Therefore, TileCal is comprised of approximately 10,000 readout
channels. The TileCal cell definition for the half of the central barrel and the extended barrel can be seen in FIG.~\ref{tilecalcells}.

The TileCal front-end electronics, located inside the outermost part of the modules, is responsible for processing the
PMT signals and transmitting them to the back-end electronics, which is responsible for calorimeter signal
acquisition.

The LHC is scheduled to undergo a major upgrade in 2024, preparing for the High Luminosity LHC operation in 2026. Therefore, the ATLAS experiment is currently going through an ambitious 10 years upgrade plan to cope with the LHC luminosity increase. Several detector components should be replaced (e.g. front-end electronics of the
calorimeters) and major changes on the trigger system are required to cope with the new luminosity
requirements.

The ATLAS upgrade plan for high luminosity is split into three phases: Phase-0 occurred during $2013-2014$ (Long Shutdown 1), and prepared the LHC for Run~2; Phase-I, foreseen for 2019 (Long Shutdown 2), will prepare the LHC for Run~3, whereafter the peak luminosity reaches $2-3 \times 10^{34}$~cm$^{2}s^{-1}$; finally, Phase-II, which is foreseen for 2024 (Long Shutdown 3), will prepare the collider for the HL-LHC operation ($5-7 \times 10^{34}$~cm$^{2}s^{-1}$).

The TileCal main activities for Phase-0 were the installation of the new low voltage power supplies, and the activation of the TileCal third layer signal for assisting the muon trigger at $1.0<|\eta|<1.3$ (TileMuon Project). During Phase-I, the replacement of the gap scintillators is foreseen. In Phase-II, a major upgrade in the TileCal readout electronics is planned. Except for the photomultipliers tubes (PMTs), most of the on- and off-detector electronics will be replaced, with the aim of digitizing all PMT pulses at the front-end level and sending them with 10~Gb/s optical links to the back-end electronics.

This work describes the major TileCal upgrade activities, focusing on the TileMuon Project and the new on- and off-detector electronics.

\section{TileCal Electronics and Upgrade Plans for High Luminosity}
\label{upgradep} 

\subsection{TileCal Electronics}

\begin{figure*}[t]
\begin{center}
\includegraphics[width=17cm]{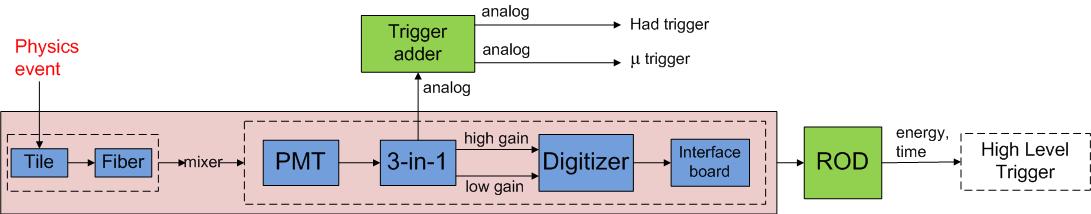}
\end{center}
\caption{\label{readoutchain}TileCal Signal Chain.}
\end{figure*}

The current TileCal signal chain can be seen in FIG.~\ref{readoutchain}. The light produced by energy depositions in the
detector is collected by optical fibers and sent to the light mixers, where several fibers are grouped together in order
to form the detector cells. The light is converted to an electrical signal in the photomultiplier tube (PMT) and is processed
by the 3-in-1 card (Front-End Board), which is responsible for signal conditioning and amplification providing three
analog signals as outputs, two for the detector readout (high and low gain) and another for triggering purpose. The low
and high gain signals are then digitized at 40~MHz by 10-bit Analog to Digital Converters (ADCs) in the Digitizer
Boards. Digital signals of all calorimeter cells in a module are merged and formatted into packages and sent via high speed optical links (Interface Board) that connect the on- and off-detector electronics. The
on-detector electronics is located in the outermost part of the TileCal module, in the electronics ``drawers''.

In the back-end electronics, the main component is the Read-Out Driver (ROD) which performs
preprocessing and gathers data coming from the front-end electronics at a maximum level 1 trigger rate of 100~kHz.
The ROD sends these data to the Read-Out Buffers (ROB) in the second level trigger \cite{trigger}.

Figure~\ref{readoutchain} also shows the signal path to the first level trigger. The TileCal first level trigger signal
is produced by analog summation of up to six signals on the Trigger Board \cite{adder} and its analog output (Had trigger) is sent to
the level 1 receiver by means of long twisted pairs cables (around 70~m). An additional output, which is a buffer to the TileCal third layer signal (D-cells), is also available for processing and is currently being prepared for use to provide coincidence for the level 1 muon trigger, envisaging fake muon rate reduction during LHC run~2 (2015-2018). The first level trigger must reduce the event rate from 40~MHz
to a maximum of 100~kHz.

\begin{figure*}[t]
\begin{center}
\includegraphics[width=17cm]{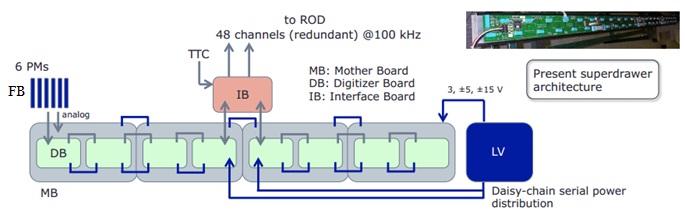}
\end{center}
\caption{\label{arch}Current architecture of the TileCal electronics drawer.}
\end{figure*}

The current architecture of the TileCal electronics drawer can be seen in FIG.~\ref{arch}. It is possible to see the low voltage distribution, the four mother boards sections (MoB), the eight digitizer boards (DiB) along the drawer and the single interface board (IB) with two optical links for redundance, sending data to the back-end electronics at a maximum rate of 100~kHz.


\subsection{TileCal Upgrade Activities}

In this section, the Tilecal major activities for Phase-0 and Phase-II will be briefly described.

\subsubsection{Phase-O Activities}

One of the main problems encountered during the TileCal operation was the occurrence of very frequent trips of the low voltage power supply (LVPS) with a very strong correlation to the integrated luminosity. In order to solve this problem, a new design of the LVPS has been developed and 38 units were tested in real conditions during 2012 data taking, showing only 1 trip in 2012. Consequently, all LVPS have been replaced by the new ones in the Long Shutdown 1 (LS1) of the ATLAS upgrade program. Additionally, a significant reduction of the electronic noise is observed and the noise distribution becomes more Gaussian.

Another important activity during LS1 was the major consolidation of the on-detector readout electronics, in particular to reinforce weak electrical connections inside the electronics drawer, the second major source of hardware failures during Run~1.

\begin{figure}[t]
\begin{center}
\includegraphics[width=8cm]{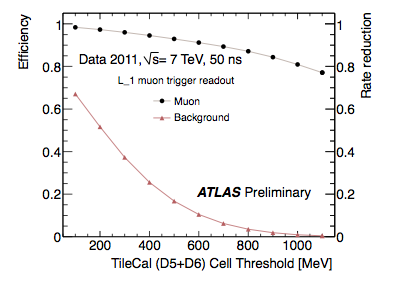}
\end{center}
\caption{\label{tmefficience} Muon detection efficiency (black dots) and background reduction (red triangles) using a prototype receiver module connected to the level 1 calorimeter trigger electronics during run~1 \cite{plot}.}
\end{figure}

In addition, during LS1 the outermost D layer cells (D5 and D6) of the TileCal extended barrel were integrated in the level 1 muon trigger, together with the end-cap muon trigger chambers, in the context of the TileMuon Project. This will reduce significantly muon fake rates originated from the slow charged particles (protons) in the region  $1.0<|\eta|<1.3$ while maintaining a good muon efficiency. Figure~\ref{tmefficience} shows the muon detection efficiency (black dots) and muon fake reduction (red triangles) as a function of the energy threshold applied to the D5+D6 TileCal cells energy, obtained with a prototype receiver module connected to the level 1 calorimeter trigger electronics during Run~1.

An intense activity was required in order to develop the system to process the analog signals from all the outer layer extended barrel cells and provide the coincidence flags for the end-cap muon trigger chambers. The TileMuon Project will be described in more details in Section~\ref{Tilemuon}.

\subsubsection{Phase-II Activities}

The main goals of the TileCal Phase-II upgrade are the replacement of the aging electronics, the increase of radiation tolerance, the improvement of system reliability (less connectors - split Main Board design mitigates against single point failure causing loss of cell), to increase data precision and to improve the level one trigger system by the availability of the full detector resolution and improved Signal to Noise Ratio (SNR).

A new on-detector electronics architecture is under design and is being incrementally tested, where three different
design options for the new TileCal front-end board are under evaluation. The increased event rate also requires larger currents in the PMT voltage divider chains.
New active dividers and a new high voltage power supply are under development.
Concerning the off-detector electronics, a new Tile PreProcessor (TilePPr) is being designed to replace the current ROD.

Along with the development of new electronics, a modification of the TileCal mechanics is being considered. The aim is to
split the present drawers into two ``mini-drawers''. This is compatible with the new electronics architecture. The
mini-drawers will simplify handling of the drawers and improve the access to the TileCal electronics since it will be
easier to open the detector to replace mini-drawers. Practical solutions to insertion and cooling and electrical
connections are being tested with different prototypes.


One upgraded drawer is currently under evaluation tests together with
aprototype of the TilePPr. This new ``Demonstrator'' drawer is backward compatible with the present system so that it can be seamlessly installed in the present system. At the end 2016, one Demonstrator drawer should be installed in the
detector in order to be tested under real conditions.

Until 2018, extensive tests should also be performed with
the three versions of the new system in test beam after which the decision about the front-end board to choose should be taken.
The production of new electronics will take place between 2020-2021. Finally, the installation is foreseen
for 2024-2026.

Section~\ref{upgradea} will describe in more details the new on-detector electronics and the Demonstrator project.

\section{TileMuon Project}
\label{Tilemuon}

The main goal of the TileMuon Project is the reduction of the muon trigger rate due to slow charged particles (protons), which interact with the muon detector in the end-cap region, by using the TileCal outermost layer signal in the extended barrel region during LHC Run~2 and Run~3. Protons emerging from the end-cap toroid and beam shielding were the main source of trigger background for the muon end-cap trigger chambers. Therefore, TileCal, which is not affected by these protons, can be used in coincidence with the muon end-cap chambers in order to reduce the muon trigger rate.

\begin{figure}[t]
\begin{center}
\includegraphics[width=8.5cm]{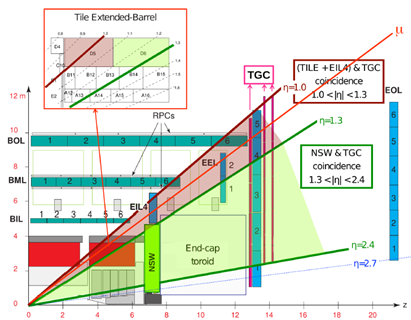}
\end{center}
\caption{\label{tilemuongeo} Illustration of the coincidence between TileCal extended barrel and the end-cap muon chamber in $1.0<|\eta|<1.3$.}
\end{figure}

Figure~\ref{tilemuongeo} illustrates a transversal cut in the ATLAS detector where is possible to verify that the muons from valid events (comes from the collision point) must cross the TileCal extended barrel before reaching the end-cap muon chambers (TGC) in the $1.0<|\eta|<1.3$ region. It is also possible to verify that this region is covered by TileCal D5 and D6 cells.

The TileCal third layer signals were already available for processing as can be seen in the FIG.~\ref{readoutchain} ($\mu$~trigger signal at the output of the trigger adder board), although they were not used during LHC Run~1. Therefore, the TileMuon Digitizer Board (TMDB) was designed to process the analog signals from all the outer layer extended barrel cells and provide the coincidence flags for the end-cap muon trigger chambers.

\subsection{TileMuon Digitizer Board}

The TMDB interfaces 8 TileCal extended barrel modules with 3 sector logic blocks from the end-cap muon chambers. A total of 16 TMDBs, installed in one VME crate, are required to fully equip the detector.

Each board receives 32 channels from the 16 TileCal cells and performs the signal digitization using 8-bit flash ADCs. The digital signals from the 32 channels feed the core FPGA (Spartan 6 from Xilinx), where the energy estimation and signal detection is performed.  The optical communication with the muon sector logic is through g-link protocol, implemented on the core FPGA, and using SFP connectors.

%
%
%

\begin{figure}[h]
\begin{center}
\includegraphics[width=6cm]{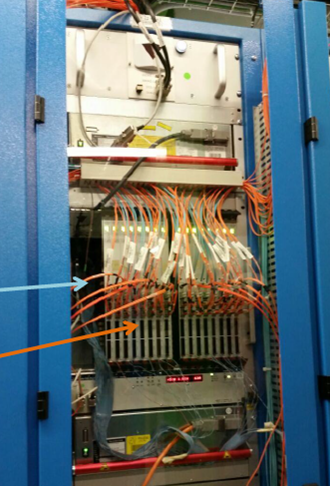}
\end{center}
\caption{\label{tmdbinstall} TileMuon installed in ATLAS USA15 cavern.}
\end{figure}

\begin{figure*}[htb]
\begin{center}
\includegraphics[width=16cm]{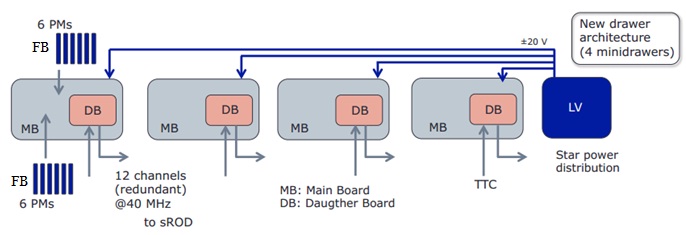}
\end{center}
\caption{\label{newarch}New drawer architecture.}
\end{figure*}

The production of the TMDB boards started by the end of 2014 and the system installation was done in the beginning of 2015. Figure~\ref{tmdbinstall} shows the system installed in ATLAS USA15 cavern. The TileMuon system is now in commissioning phase preparing for the beginning of operation late in 2015.

\section{New Front-End Electronics for Phase-II}
\label{upgradea}


The foreseen architecture for the new on-detector electronics can be seen in FIG.~\ref{newarch}. The new readout
electronics is composed of: new Front-End boards (FE) that provide conditioning for the PMT signals as well
amplification, digitization (depending on the Front-End option) and calibration functionalities; new Main Boards (MB)
providing digitization (depending on the Front-End board) and control; and new Daughter Boards (DB) which provide data
processing and interface with the back-end electronics via optical links at up to 40~GHz rate with redundance. The star power distribution along with local point-of-load voltage regulators in the new drawer reduces the voltage deviations and the noise coupling along the drawer. Another
important change on the drawer architecture is the replacement of a single interface board (actual design) by four
Daughter Boards, improving the system robustness and decoupling the drawer electronics into independent units.

\subsection{Front-End - Modified 3-in-1 Card}

The Enrico Fermi Institute (University of Chicago) is developing a modified version of the present 3-in-1 card
\cite{3in1}. This Front-End Board is composed of discrete components and can be divided in three stages: the fast
signal processing chain, the slow signal processing chain and the calibration electronics and the control bus
interface.

The fast signal processing chain includes a 7-pole passive LC shaper, bi-gain clamping amplifiers with a gain ratio of
32 and a pair of differential drivers feeding the analog signals from the low-gain channel and the high-gain channel to
the ADCs which are placed in the Main Board. The slow signal processing chain includes a programmable 3-gain integrator
which monitors the PMT current induced by a Cesium source and the minimum bias current induced during the collisions.
Finally, the last stage includes a precise charge injection circuit, integrator gain control and the control bus
interface. This modified version has better linearity and a lower noise level than the previous version. The
prototype of the Modified 3-in-1 Card has been built using COTS components and has passed initial radiation tests.

\begin{figure}
\begin{center}
\includegraphics[width=6cm]{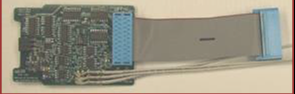}
\end{center}
\caption{\label{3in1} Modified 3-in-1 Card.}
\end{figure}

The Modified 3-in-1 Card is currently equipping the Demonstrator drawer and due to that a different version of the board was produced, including an analog output for the Trigger Board in order to preserve the system compatibility with the current level 1 trigger system. Figure~\ref{3in1} shows the Modified 3-in-1 Card.

\subsection{Front-End - QIE Chip}

The Argonne National Laboratory together with Fermilab are working on the design of a front-end board which includes a new version of the Charge
(Q) Integrator and Encoder (QIE) chip developed in collaboration with Fermilab and CMS HCAL. The QIE includes
a current splitter composed of 23 splitter transistors, providing 4 different ranges (16/23, 4/23, 2/23, 1/23), followed
by a gated integrator and an on-board 7-bit flash ADC to cover a dynamic range of 18 bits. In this way, only a simple
digital interface is needed to communicate with the Main Board. The QIE also includes a charge injection circuit for
calibration and an integrator for source calibration. The QIE does not perform pulse shaping, minimizing pile up problems and allowing raw PMT pulses to be measured.

\begin{figure}
\begin{center}
\subfloat[QIE chip.]{\includegraphics[width=2cm]{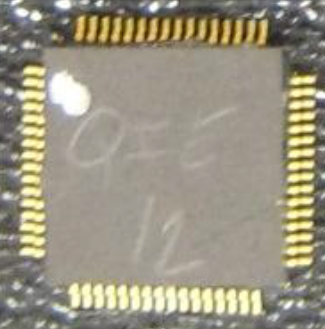}}\\
\subfloat[QIE prototype test board.]{\includegraphics[width=6cm]{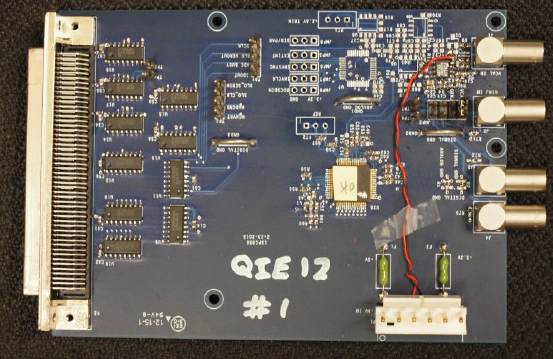}}
\end{center}
\caption{\label{QIE} QIE chip and its prototype test board.}
\end{figure}

At the present time several successful tests were performed, showing that the performance and radiation tolerance meet the TileCal requirements. The second version of the chip (QIE12) is already in hand and is scheduled for test beam in 2016. Figure~\ref{QIE} shows the QIE12 chip and its prototype test board.

\subsection{Front-End - FATALIC ASIC}

FATALIC means Front-End for Atlas TileCal Integrated Circuit \cite{fatalic}, which is being designed at Laboratoire de
Physique Corpusculaire in Clermont-Ferrand (LPC). FATALIC includes a multi-gain current conveyor (CC) with three
different gains (1, 8, 64) which cover the full dynamic range of the PMT signal, followed by a shaper in order to
improve the SNR. The readout chain is completed using an external 12-bit pipelined ADC with a
sampling rate of 40~MHz also developed at LPC and called Twelve bits ADC for ATLAS TileCal Integrated Circuit
(TACTIC). Moreover FATALIC includes an integrator and a 10-bit ADC with a low sampling rate for calibration purposes.
Both chips are designed using the CMOS IBM 130~nm technology.

\begin{figure}
\begin{center}
\includegraphics[width=6cm]{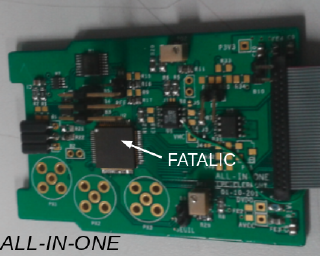}
\end{center}
\caption{\label{fatalic} All-In-One Board for the FATALIC.}
\end{figure}

The prototype FATALIC version 4 is currently under tests, showing promising results. The first tests revealed a very good signal to noise ratio and a good linearity over the three gains. The front-end board for the FATALIC is called All-In-One Board and can be seen in FIG.~\ref{fatalic}.

\subsection{Main Board and Daughter Board}

\begin{figure}
\begin{center}
\includegraphics[width=8.5cm]{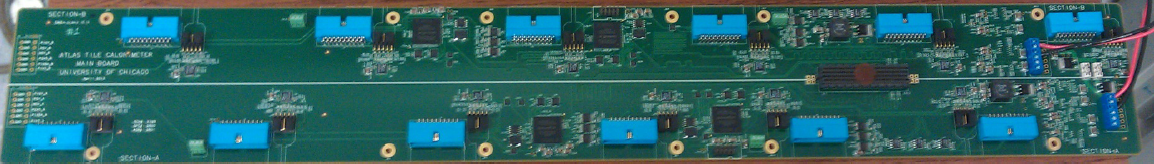}
\end{center}
\caption{\label{MB} Main Boards version 2. The line in the middle of the board indicate the board symmetry for redundancy.}
\end{figure}

The Main Board \cite{mbdb,demons1} is responsible for the digital control of the FEs, data organization and for the
transmission of the data to the Daughter Board. The current prototype design (version 2) digitizes the signals coming from 12 Modified 3-in-1 Cards by using twenty four 12-bit ADCs working at a sampling rate of 40~MHz and is being used in the Demonstrator. Figure~\ref{MB} shows the Main Board version 2.

\begin{figure}
\begin{center}
\includegraphics[width=8cm]{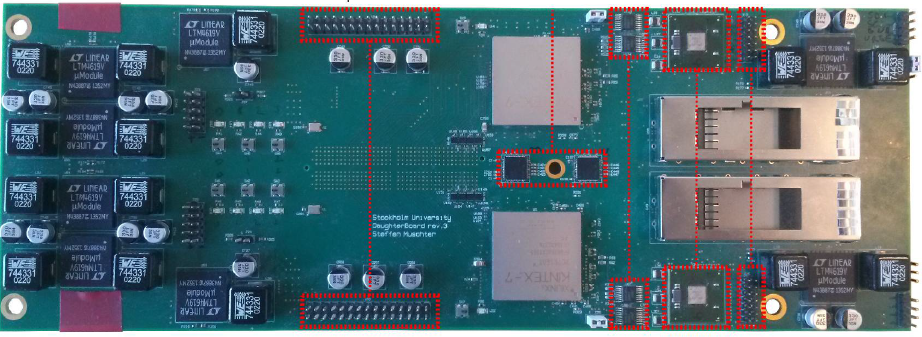}
\end{center}
\caption{\label{DB} Daughter Board version 3. It can be seen that the bottom is a mirror of the top part for redundance.}
\end{figure}

The Daughter Board \cite{mbdb,demons} is intended to serve as a processing board in the next TileCal electronics drawer and is designed for redundancy and high data throughput readout. The two separately programmable FPGAs are responsible for reading out signals originating from the same tile cells but from different sides of the scintillating tiles. This means that they process equivalent data. If one chain fails it can be replaced by the other (there is a loss of statistics though). The Daughter Board sends the digitized data to the super Read Out Driver (sROD) via high-speed links using the GBT
protocol \cite{gbt}. In order to perform these functions, the third version of the Daughter Board includes two Xilinx Kintex 7 FPGAs, one $4\times10$~Gb/s QSFP+ and two transmitters using two QSFPs connectors. The board includes fully redundance. The Daughter Board version 3 can be seen in FIG.~\ref{DB}.

\begin{figure}[]
\begin{center}
\includegraphics[width=9cm]{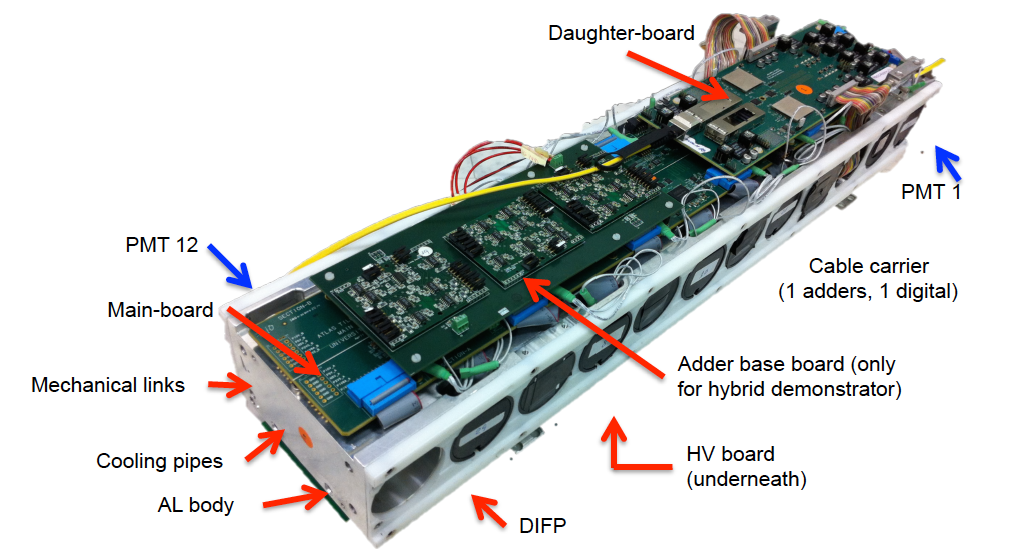}
\end{center}
\caption{\label{demonstrator} Demonstrator drawer. The modified 3-in-1 boards are placed inside the PMTs in the holes of the aluminum body, the Daughter Board (DB) is on the top right, the Main Board is below DB and the Adder Base Board. The adders are mounted on the Adder Base Board in order to provide the analog trigger signal.}
\end{figure}

The Daughter and the Main boards have been designed by the Stockholm University and the Enrico Fermi Institute
(University of Chicago). The Main Board version 2 and the Daughter Board version 3 have been produced and tested. Both boards are being used on the Demonstrator drawer \cite{demons,demons1}.


\subsection{Demonstrator}

A Demonstrator project has been established in order to test the feasibility and performance of the proposed upgrade electronics. It should contain as much of the final Phase 2 design as possible while being compatible with the  present system.

The tests with a hybrid demonstrator drawer, which provides both digital and analog trigger data, are currently being performed. Test beam campaigns are planned for 2015-2016 before the installation in the detector by the end of 2016.

Figure~\ref{demonstrator} shows one demonstrator drawer that is currently being tested at CERN. The Modified 3-in-1 Cards are placed inside the PMTs in the holes of the aluminum body, the Daughter Board (DB) is on the top right of the figure, the Main Board is below DB and the Adder Base Board. The adders are mounted on the Adder Base Board in order to provide the analog trigger signal to level~1.

%
%
%
%
%

\section{Conclusions}
\label{conclusions}

TileCal is facing challenging R\&D activities due to the required upgrade for the luminosity increase at LHC.

For the LHC Run~2, the main upgrade activities were related to the replacement of the LVPS and the activation of the D-layer signal in order to be used in coincidence with the muon end-cap trigger.

The replacement of the LVPS improved the detector reliability and, in addition, a significant reduction of the electronic noise was observed.

In order to receive and process the TileCal muon trigger signals in the extended barrel region, the TMDB was designed, fully tested and produced. The TileMuon system is now in commissioning phase preparing for the beginning of operation by the end of 2015.

Concerning the upgrade Phase-II, the on- and off-detector electronics must be
redesigned. For the on-detector electronics, three different Front-End boards approaches are being considered.
Additionally, in order to provide sufficient data processing, control and interface with the new back-end electronics, a Main Board and Daughter Board combination are being designed.

Extensive laboratory tests are currently being
performed with one full Demonstrator drawer, providing promising results.

One Demonstrator drawer should be installed into the detector by the end of 2016. The final decision about the
Front-End Board design should be taken by the end of 2018. The production of the new Tilecal readout electronics will
take place during 2019-2020 and should be prepared for installation during Phase-II.

\begin{acknowledgments}
I would like to thank my colleagues from the Tile Calorimeter Group for the fruitful discussions and UFJF, CNPq, CAPES and FAPEMIG from Brazil for the support to this work.
\end{acknowledgments}



\bibliography{TileUpgradeLISHEP}

\end{document}